\documentclass[superscriptaddress,prb,twocolumn,aps,showpacs]{revtex4}

\usepackage{graphicx}
\usepackage{dcolumn}
\usepackage{bm}

\begin{document}
\title{Observation of Jonscher Law in AC Hopping Conduction of Electron-Doped Nanoporous Crystal 12CaO$\cdot$7Al$_{2}$O$_{3}$ in THz Frequency Range}

\author{H. Harimochi}
\affiliation{Department of Quantum Matter, ADSM, Hiroshima University, 1-3-1 Kagamiyama, Higashi-Hiroshima 739-8530, Japan}
\author{J. Kitagawa \footnote{jkita@hiroshima-u.ac.jp}}
\affiliation{Department of Quantum Matter, ADSM, Hiroshima University, 1-3-1 Kagamiyama, Higashi-Hiroshima 739-8530, Japan}
\author{M. Ishizaka}
\affiliation{Department of Quantum Matter, ADSM, Hiroshima University, 1-3-1 Kagamiyama, Higashi-Hiroshima 739-8530, Japan}
\author{Y. Kadoya}
\affiliation{Department of Quantum Matter, ADSM, Hiroshima University, 1-3-1 Kagamiyama, Higashi-Hiroshima 739-8530, Japan}
\author{M. Yamanishi}
\affiliation{Department of Quantum Matter, ADSM, Hiroshima University, 1-3-1 Kagamiyama, Higashi-Hiroshima 739-8530, Japan}
\author{S. Matsuishi}
\affiliation{Materials and Structures Laboratory, Tokyo Institute of Technology, Nagatsuta Midori-ku, Yokohama 226-8503, Japan}
\author{H. Hosono}
\affiliation{Materials and Structures Laboratory, Tokyo Institute of Technology, Nagatsuta Midori-ku, Yokohama 226-8503, Japan}

\date{}

\begin{abstract}
We have performed terahertz time-domain spectroscopy of carrier-doped nanoporous crystal 12CaO$\cdot$7Al$_{2}$O$_{3}$ showing the Mott variable range hopping at room temperature. The real part of the dielectric constant clearly demonstrates the nature of localized carriers. The frequency dependence of both the real and imaginary parts of the dielectric constant can be simply explained by assuming two contributions: a dielectric response by the parent compound with no carriers and an AC hopping conduction with the Jonscher law generally reported up to GHz range. The possible obedience to the Jonscher law in the THz range suggests a relaxation time of the hopping carriers much faster than 1ps in the carrier-doped 12CaO$\cdot$7Al$_{2}$O$_{3}$.
\end{abstract}

\maketitle

\clearpage


Terahertz time-domain spectroscopy (THz-TDS) is now bridging a spectroscopic technology gap between an optical method represented by Fourier transform infrared spectroscopy and an electrical one like an impedance spectroscopy.
A rapid rise in attractiveness of THz-TDS would be mainly due to the fact that both the real and imaginary parts of optical constants can be extracted without using the Kramers-Kronig transformation in addition to the nondestructive and noncontact features.
THz-TDS has highly succeeded in demonstrating that complex electrical conductivities, $\sigma^{*}$'s, of materials such as a doped Si and a doped GaAs follow the Drude model with an occasional modification\cite{Exter,Katzenellenbogen,Jeon}.

Several intriguing physical phenomena exist in disordered matter in the so-called THz gap.
For example the Boson peak interpreted as a kind of low energy excitation has already been investigated by THz-TDS\cite{Kojima}.
Another significant subject is $\sigma^{*}$ in THz range.
A hopping conduction is frequently observed in disordered matter like an amorphous semiconductor or an ionic conductor.
It is well known that the real part of $\sigma^{*}$, $\sigma_{1}$, of the hopping conduction shows a sublinear frequency dependence varying as $\omega^{s}$ (0$<s<$1) in the frequency range below 1GHz\cite{Jonscher}, which is called the Jonscher power law.
On the other hand, a contribution of a phonon vibration becomes dominant in the spectral range above an infrared frequency.
Recently it has been suggested that a deviation from the Jonscher power law commonly occurs among disordered matter in the THz gap\cite{Lunkenheimer}, in which $\sigma_{1}$ exhibits a superlinear frequency dependence.
The conclusion is chiefly based on a frequency-domain spectroscopic experiment and a discussion with neglect of a possible phonon contribution in the THz gap.
Thus the above mentioned situation motivated us to employ THz-TDS for an ideal disordered material showing a hopping conduction in order to obtain a more deep insight in $\sigma^{*}$ in the THz gap.

In this study we focused on nanoporous crystal 12CaO$\cdot$7Al$_{2}$O$_{3}$ (C12A7).
The crystal structure of C12A7 is composed of sub-nanometer-sized cages with an inner diameter 0.44nm, and each cage is coordinated by 12 cages to form a 3-dimensional cubic lattice sharing an open mouth with a neighboring cage\cite{Hayashi,Matsuishi}.
There are 3 inequivalent crystallographic sites for the oxygen\cite{Bartl}.
One of the sites is located in the center of the cage, and 1/6 of the site is statistically occupied by an oxygen ion O$^{2-}$ in the stoichiometric state.
This oxygen characterized by loosely bounded state is called "free oxygen ions" the concentration of which is about 1$\times$10$^{21}$cm$^{-3}$.
C12A7 itself is a wide-gap insulator.
Recently, it was reported that C12A7 can be converted to an $n$-type conductor by replacing the free oxygens with electrons using chemical reactions with Ca metals\cite{Matsuishi}.
The extent of this substitution is controllable by tuning the reaction conditions.
Then, the sample changes gradually from insulator to degenerate semiconductor depending on the carrier electron concentrations, $n_{c}$.
Within a range, 2$\times$10$^{18}$$\sim$2$\times$10$^{20}$cm$^{-3}$, of $n_{c}$, C12A7 shows the Mott variable range hopping even at room temperature\cite{Matsuishi}.
The highest $n_{c}$ of samples measured in this study is 5$\times$10$^{18}$cm$^{-3}$.
Therefore even in this sample obeying the Mott variable range hopping, only 0.25\% of the free oxygen ions is replaced with electrons.
This means that there would be little difference in $\sigma^{*}$ arising from another contribution except for the hopping carriers between the carrier-doped C12A7 and the parent compound (C12A7-P) with no carriers, and will lead to an exclusive conclusion about $\sigma^{*}$ purely stemming from the hopping carriers.
In this article, we report a THz-TDS measurement of the carrier-doped C12A7 and C12A7-P.


We have prepared a C12A7-P and two carrier-doped C12A7 single crystalline samples.
The $n_{c}$'s are 3$\times$10$^{17}$cm$^{-3}$ (C12A7-L) and 5$\times$10$^{18}$cm$^{-3}$ (C12A7-H), respectively.
The detailed procedure of the sample preparation is reported in ref.8.
All samples were cut into platelets with thickness 995$\mu$m for C12A7-P, 355$\mu$m for C12A7-H and 340$\mu$m for C12A7-L, respectively.
A transmission THz-TDS was performed with a system schematically depicted in Fig.1.
Freely propagating THz pulses were generated by an InAs semiconductor surface optically excited by a mode-locked Ti:sapphire laser.
This laser produces ultrafast pulses with a central wavelength of 800nm and a duration of 150fs at a pulse repetition rate of 76MHz.
A time evolution of THz pulses transmitted through a sample was then detected by a low-temperature-grown (LT) GaAs photo-conductive (PC) antenna triggered by time-delayed pulses of the laser.
To eliminate the effect of moisture, an optical path of THz pulses was enclosed by a box purged with N$_{2}$ gas.
Our THz-TDS system has a frequency bandwidth between 0.3 and 3THz.
For C12A7-H, a low-temperature THz-TDS was performed down to 70K by cooling the sample in a commercial cryostat. 

\begin{figure}[hbtp]
\begin{center}
\includegraphics[width=\linewidth]{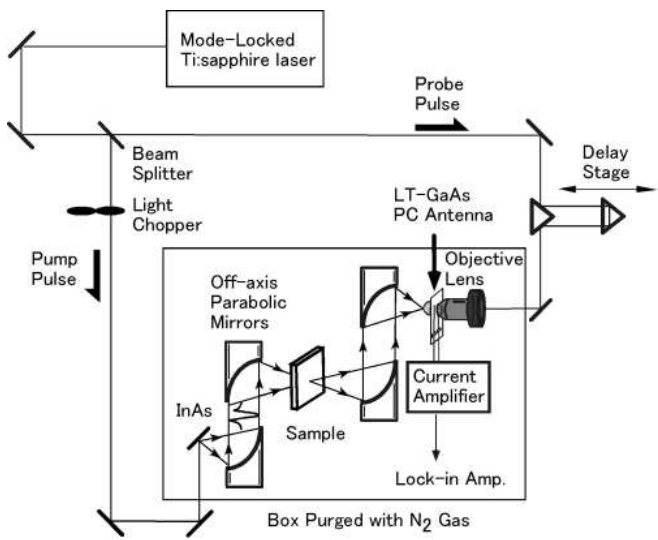}
\caption{The schematic view of the THz-TDS measurement system.}
\label{fig1}
\end{center}
\end{figure}

Figure 2 shows measured THz pulses without sample (denoted as REF) and transmitted through C12A7-P, -L and -H at room temperature, respectively.
The attenuated amplitude of C12A7-H compared to that of C12A7-L would be mainly due to a large carrier absorption in C12A7-H.
The fact that the C12A7-P is thicker than C12A7-H and -L causes the retardation of the THz pulses for C12A7-P.
Another C12A7-P sample with a thickness $<$ 995 $\mu$m displays less attenuated THz pulses (not shown) than does the present one, indicating that C12A7-P itself possesses sources of the absorption in the THz range.
The frequency dependences of amplitude transmittances and their phases of C12A7 were obtained by the Fourier transformation of the THz pulses in Fig.2.
Comparing them with those of REF, complex dielectric constants, $\epsilon^{*}(=\epsilon_{1}+i\epsilon_{2})$'s, were obtained as shown in Fig.3.
Strong reductions of the transmittances above about 2THz in all samples make the effective bandwidth narrower than that of the measurement system itself.
There is little difference between $\epsilon^{*}$ of C12A7-P and that of C12A7-L.
The rather small differences at low and high frequency regimes may be due to an existence of a very weak dielectric response of carriers in C12A7-L and/or a growth of an experimental error.
In contrast, $\epsilon^{*}$ of C12A7-H is much different from C12A7-P.
In particular, $\epsilon_{2}$ decreases with increasing frequency and exhibits a rather shallow minimum around 1.4THz.

\begin{figure}[hbtp]
\begin{center}
\includegraphics[width=\linewidth]{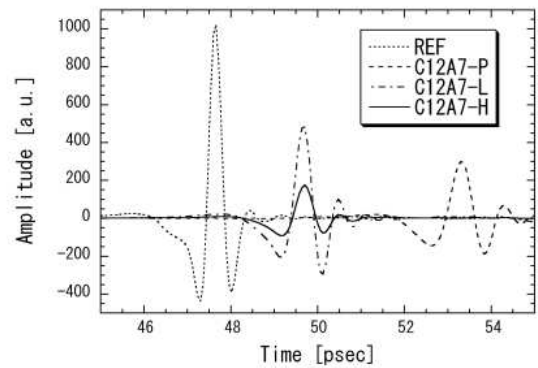}
\caption{The time evolution of the THz pulses without sample (REF) and transmitted through C12A7-P, -L and -H at room temperature.}
\label{fig2}
\end{center}
\end{figure}

For all samples including C12A7-P with no carriers, the finite value of the $\epsilon_{2}$ strongly suggests some kinds of absorptions in the THz range.
At first $\epsilon^{*}$ of C12A7-P is analyzed with a Lorentz model because THz phonon vibration modes are the most plausible sources for the dispersive $\epsilon^{*}$.
It is provided that the frequency dependence of $\epsilon^{*}$ of C12A7-P can be described by the Lorentz model with a single resonance as follows:
\begin{equation}
\epsilon^{*}(\omega)=\epsilon_{\infty}+\frac{(\epsilon_{0}-\epsilon_{\infty})\omega^{2}_{0}}{\omega_{0}^{2}-\omega^{2}-i\omega\gamma},
\label{equ:Lorentz}
\end{equation}
where $\omega_{0}$ represents a resonance frequency, $\gamma$ a damping constant, and $\epsilon_{\infty}$ and $\epsilon_{0}$ dielectric constants at the frequency much higher and lower than $\omega_{0}$, respectively.
In Fig.3, the dash-dotted curves are calculated $\epsilon^{*}$ with $\omega_{0}$=1.59$\times$10$^{13}$Hz, $\gamma$=4.94$\times$10$^{12}$Hz, $\epsilon_{\infty}$=6.03 and $\epsilon_{0}$=7.04 describing the experimental data very well.
The $\omega_{0}$ means an existence of a resonance absorption around 2.5THz, for which a spectroscopic measurement above 2THz is desired.

No conspicuous difference between $\epsilon^{*}$ of C12A7-P and that of C12A7-L urges negligible carrier contributions in $\epsilon^{*}$ of C12A7-L.
The conjecture is consistent with the fact that the proportion of the replacement of the free oxygen ions with electrons is only about 0.02\% in C12A7-L.
A dc electrical conductivity, $\sigma_{dc}$, of C12A7-L is estimated to be 0.015$\Omega^{-1}$cm$^{-1}$ by using $\sigma_{dc}$ of 100$\Omega^{-1}$cm$^{-1}$ at $n_{c}$=2$\times$10$^{21}$cm$^{-3}$ reported in ref.8 on the assumption that the mobility is independent of $n_{c}$.
The value corresponds to that of the sample "c" displaying no Mott variable range hopping but an Arrhenius-type conductivity at room temperature in ref.8.
Furthermore $\epsilon^{*}$ of C12A7-L can be roughly estimated by an extrapolation of experimental $\epsilon^{*}$ of C12A7-H assuming that increment from $\epsilon^{*}$ of C12A7-P is proportional to $n_{c}$.
The estimated $\epsilon^{*}$ is almost the same as experimental one.
Therefore no carrier effect on the dielectric response of C12A7-L would be due to the extremely low $n_{c}$ and/or the transport mechanism which does not work well in the THz range.
On the other hand, $\sigma_{dc}$ of C12A7-H is estimated to be 0.25$\Omega^{-1}$cm$^{-1}$ by the above mentioned method.
The value corresponds to that of the sample "d" following the Mott variable range hopping in ref.8.
Even in this sample the replacement of only about 0.25 \% of the free oxygen ions occurs, implying that $\epsilon^{*}$ of C12A7-P are responsible for that with no carrier contribution in C12A7-H.
The difference between $\epsilon^{*}$ of C12A7-H and that of C12A7-P can be thus accurately ascribed to a dielectric response of carriers in C12A7-H.

It is important to know whether the carriers in C12A7-H are localized or itinerant in the THz range, on which $\epsilon_{1}$ especially informs us\cite{Pimenov}.
According to the Drude model sufficiently being able to handle itinerant carriers, $\epsilon_{1}(\omega)$ of carrier-doped sample is expressed as follows:
\begin{equation}
\epsilon_{1}(\omega)=\epsilon_{1:P}(\omega)-\frac{\sigma_{2}}{\omega\epsilon_{0}},
\label{equ:Drudeepsilon}
\end{equation}
where $\epsilon_{1:P}$ is $\epsilon_{1}$ of the parent compound, $\epsilon_{0}$ the vacuum permittivity and $\sigma_{2}$ the imaginary part of $\sigma^{*}$ caused by the itinerant carriers.
Eq.(\ref{equ:Drudeepsilon}) demonstrates the relation that $\epsilon_{1}$ is reduced from $\epsilon_{1:P}$.
This is the proper consequence judging from the fact that the negative $\epsilon_{1}$ (the second term in Eq.(\ref{equ:Drudeepsilon})) is characteristic with ordinary metals below the plasma frequency.
An opposite relation is suggested by hopping conduction mechanisms including the Mott variable range hopping, that is\cite{Elliott}
\begin{equation}
\epsilon_{1}(\omega)=\epsilon_{1:P}(\omega)+\frac{\sigma_{2}}{\omega\epsilon_{0}}.
\label{equ:Hoppingepsilonreal}
\end{equation}
Eq.(\ref{equ:Hoppingepsilonreal}) is derived by treating the hopping carrier transport as the displacement current which leads to the positive $\epsilon_{1}$(the second term in Eq.(\ref{equ:Hoppingepsilonreal})).
As can be easily seen from Fig.3(a), $\epsilon_{1}$ of C12A7-H larger than that of C12A7-P implies that the nature of localization is realized presumably due to maintenance of the hopping conduction in the THz range.

\begin{figure}[hbtp]
\begin{center}
\includegraphics[width=\linewidth]{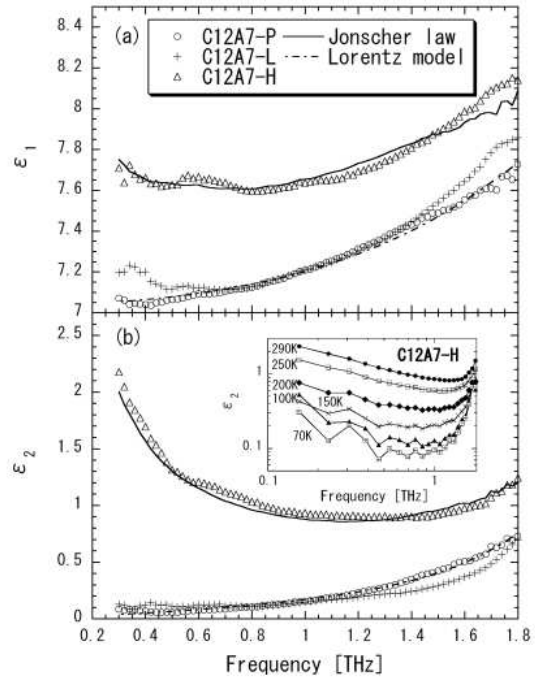}
\caption{The frequency dependence of $\epsilon^{*}$ of C12A7-P ($\circ$), -L ($+$) and -H ($\triangle$) at room temperature. The solid curves are $\epsilon^{*}$ calculated by using the Jonscher law. The dash-dotted curves represent $\epsilon^{*}$ calculated by the Lorentz model with the single resonance. The inset of (b) is the temperature dependence of $\epsilon_{2}$ of C12A7-H.}
\label{fig3}
\end{center}
\end{figure}

The evidence of the nature of localized carriers in the THz range puts forward a progressive discussion on the power-law frequency dependence of $\sigma_{1}$ as mentioned in the introduction.
It should be noted here that THz-TDS gives both $\epsilon_{1}$ and $\epsilon_{2}$ separately with each other.
Furthermore they originally fulfill the Kramers-Kronig relation, the nature of which requires a broadband physical information.
Therefore a successful simultaneous fitting of $\epsilon_{1}$ and $\epsilon_{2}$ would provide a rather reliable decision about the power law dependence even if the frequency range is narrow.
We have assumed that $s$ is frequency independent and used the semi-empirical $\sigma(\omega)$ formula of the Jonscher law\cite{Jonscher,Elliott} fulfilling the Kramers-Kronig relation given as 
\begin{equation}
\sigma(\omega)=\sigma_{dc}+A\omega^{s}+iA\omega^{s}\tan(\frac{s\pi}{2}),
\label{equ:sigma}
\end{equation}
with which $\epsilon_{1}$ and $\epsilon_{2}$ are calculated using Eq.(\ref{equ:Hoppingepsilonreal}) and $\epsilon_{2}(\omega)=\epsilon_{2:P}(\omega)+\sigma_{1}(\omega)/(\omega\epsilon_{0})$.
The $\sigma_{dc}$ is fixed to be 0.25$\Omega^{-1}$cm$^{-1}$ estimated in the above discussion.
The solid ($\sigma_{dc}$=0.25$\Omega^{-1}$cm$^{-1}$, A=7.3$\times$10$^{-10}$$\Omega^{-1}$cm$^{-1}$s$^{-0.65}$, $s$=0.65) curves in Figs.3(a) and (b) are calculated $\epsilon^{*}$ and they agree well with experimental ones.
A deviation in the fitting would be due to an oversimplification of the semi-empirical formula as mentioned below.
To further investigate $\sigma^{*}$ in the THz range, the low-temperature THz-TDS measurement has been performed for C12A7-H.
The experimental results of $\epsilon_{2}$ containing a contribution from the parent sample is shown in the inset of Fig.3(b).
At 70K, $\sigma_{dc}$ is about 10$^{-6}$$\Omega^{-1}$cm$^{-1}$ which is 5 orders of magnitude smaller than that at room temperature.
If there is an underlying superliear power law at room temperature, this becomes dominant with decreasing temperature and a growth of $\epsilon_{2}$ with increasing frequency would be expected since $\epsilon_{2}$ has a term of $\sigma_{1}(\omega)/(\omega\epsilon_{0})\propto\omega^{s-1}$.
Although the experimental results are consistent with the sublinear power law at least up to 1.2THz, the nearly constant $\epsilon_{2}$ at low temperatures below 1THz may be the precursor of a superlinear power law and such an alternative interpretation for $\sigma^{*}$ is also possible at the present stage.
The possible obedience to the Jonscher law indicates that a hopping transport is realized even in the THz range, which means that carriers can move via a hopping or a tunneling process in response to the THz electric fields.
This suggests a relaxation time of hopping carriers much faster than 1ps.

In the present discussion, the frequency dependence of $s$ is not taken into account.
Indeed only when $s$ is assumed to be constant, the oversimplified Eq.(\ref{equ:sigma}) can be derived.
Several models are proposed to explain an AC hopping conduction up to GHz range.
Each model predicts unique frequency and temperature dependences on $s$\cite{Elliott}.
The fact that $\epsilon^{*}$ of C12A7-H can be explained by means of Eq.(\ref{equ:sigma}) indicates a very weak frequency dependence on $s$.
But a detailed analysis of the temperature dependence of $s$ would assert the applicability of the specific model or the requirement of constructing a new model.

In summary we have employed THz-TDS for the first time to shed light on the dielectric response of hopping carriers of disordered materials in the THz gap to the best of our knowledge.
The carrier-doped C12A7 sample reported in this article shows the Mott variable range hopping.
The nature of localization due to the hopping conduction is inferred from $\epsilon_{1}$ in the THz range.
The accurate components of the hopping carriers in both $\epsilon_{1}$ and $\epsilon_{2}$ can be extracted and they are well described by simply combining $\epsilon^{*}$ of the parent compound and that of hopping carriers which is deduced from the Jonscher law generally observed up to GHz range.
But the experimental results may be also interpreted by an underlying superlinear power law and further study is needed. 
The possible obedience to the Jonscher law in the THz range suggests a relaxation time of hopping carriers much faster than 1ps in the carrier-doped C12A7.

This work was supported by the Foundation for C\&C Promotion and a Grant-in-Aid for Young Scientists B (No.16760013), Scientific Research C (No.15560032) and  Creative Scientific Research (No.16GS0205) of MEXT and for Strategic Information and Communications R\&D Promotion Programme of Ministry of Public Management, Home Affairs, Posts and Telecommunications, Japan.

\end{document}